\documentclass[10pt,conference,letterpaper]{IEEEtran}
\pdfoutput=1
\usepackage{times,amsmath,epsfig}
\usepackage{amsfonts}
\usepackage{graphicx}
\usepackage{subfig}
\usepackage{alltt}
\usepackage[utf8]{inputenc}
\usepackage{algorithmicx}
\usepackage{algpseudocode}
\usepackage{cite}

\usepackage{booktabs}

\algnotext{EndIf}
\algnotext{EndWhile}
\algnotext{EndFor}

\title{Distributed Community Detection with the WCC Metric}
\author{%
{Matthew Saltz{\small $~^{1}$}, Arnau Prat-P\a'erez{\small $~^{2}$}, David Dominguez-Sal{\small $~^{3}$} }%
\vspace{1.6mm}\\
\fontsize{10}{10}\selectfont\itshape
$^{}$\,Computer Architecture Department\\
Universitat Polit\a'ecnica de Catalunya\\
Barcelona, Spain, 08034\\
\fontsize{9}{9}\selectfont\ttfamily\upshape
$^{1}$\,msaltz@ac.upc.edu\\
$^{2}$\,aprat@ac.upc.edu\\
$^{3}$\,ddomings@ac.upc.edu%
\vspace{1.2mm}\\
\fontsize{10}{10}\selectfont\rmfamily\itshape
}
\begin{document}
\maketitle
\begin{abstract} 
Community detection has become an extremely active area of research in recent
years, with researchers proposing various new metrics and algorithms to address
the problem. Recently, the Weighted Community Clustering (WCC) metric 
was proposed as a novel way to judge
the quality of a community partitioning based on the distribution of triangles
in the graph, and was demonstrated to yield superior results over other commonly
used metrics like modularity. The same authors later presented a parallel
algorithm for optimizing WCC on large graphs. In this paper, we propose a new
distributed, vertex-centric algorithm for community detection using the WCC
metric. Results are presented that demonstrate the algorithm's performance and
scalability on up to 32 worker machines and real graphs of up to 1.8 billion
vertices. The algorithm scales best with the largest graphs, and to our knowledge,
it is the first distributed algorithm for optimizing the WCC metric.
\end{abstract}

\section{Introduction}
\label{sec:intro}

Due to the generality of the graph as a data structure, graphs correspond well
to many different systems in the real world, like social networks, molecules,
road maps, and more; and many problems can be expressed intuitively and solved
using a graph representation.  One such problem whose solution has many
applications is that of \emph{community detection} -- automatically identifying
groups of vertices that are tightly connected among themselves and loosely
connected with the rest of the graph. In social networks, for example, the
identification of communities can help with targeted marketing; or in a network
of items that are frequently purchased together, community detection could be
used to make recommendations. 

As the graphs being operated on become larger and larger, the ability to process
them in memory on one machine becomes infeasible due to both time and memory
constraints.  For these two reasons, complexity and size, distributed algorithms
have become necessary to solve problems on large graphs.  In this paper, we
present a distributed algorithm for optimizing WCC \cite{wcc}, a recently proposed metric for
judging the quality of community partitionings. The algorithm  scales well on
real graphs of up to 1.8 billion edges and outperforms a parallel, centralized
algorithm that also seeks to optimize WCC \cite{scd}. The algorithm follows the
vertex-centric paradigm introduced by the Pregel platform \cite{pregel}, and to
the best of our knowledge, it is the first distributed algorithm for optimizing
the WCC metric.

The structure of the paper is as follows. 
In Section~\ref{sec:relwork}, we begin by presenting an overview of related work
in community detection and distributed community detection.  
Next, in
Section~\ref{sec:background}, we introduce background material and the
terminology used in the rest of the paper.  Following this, the proposed
algorithm is explained in Section~\ref{sec:algorithm}, followed by
experimentation in Section~\ref{sec:experimentation}. We conclude with a
discussion of future work.

\section{Related Work}
\label{sec:relwork}
Most of the research on community detection algorithms has focused on
single threaded algorithms on SMP machines. The list of proposals is rich and
diverse, with those based on modularity maximization forming the most prominent family
of community detection algorithms~\cite{newman2004finding}.  Modularity is a
community detection metric that rewards those partitions with communities with
an internal edge density larger than that expected in a null model. Several
strategies have been proposed for its optimization, such as agglomerative
greedy~\cite{clauset2004finding} or simulated
annealing~\cite{medus2005detection}. One of the most famous and widely used
community detection algorithms based on modularity maximization is the
\emph{Louvain} method~\cite{blondel2008fast}, a multilevel approach that
scales to graphs with hundreds of millions of objects. However, the quality of
its results decreases considerably as the size of the graph
increases~\cite{lancichinetti2009community}. More importantly, it has been
reported that modularity has resolution
limits~\cite{fortunato2007resolution,bagrow2012communities}, which means that
modularity is unable to detect small and well-defined communities when the
graph is large. Related to this, recent studies have proven not only that
modularity has detectability issues~\cite{nadakuditi2012graph} (i.e. it is not
able to identify communities even if they are well defined), but also that the
identification of well-defined communities is more difficult than 
ill-defined ones~\cite{radicchi2014paradox}. Although it has not been studied
whether or not $WCC$ also suffers from these problems, properties presented
in ~\cite{wcc} suggest that algorithms based on $WCC$ are able to deliver
cohesive and structured communities regardless of the size of the graph.

There also exist several proposals based on random walks.  The intuition is
that in a random walk, the probability of remaining inside of a community is
higher than going outside, due to the higher density of internal edges. This
strategy is the main idea exploited in Walktrap~\cite{Pons06}. Another
algorithm based on random walks that is highly adopted in the literature is
Infomap~\cite{rosvall2008maps}, which searches for a codification for
describing random walks based on communities. The codification that requires
the least amount of memory (attains the highest compression rates) is selected.
According to the comparison performed by Lancichinetti et
al.~\cite{lancichinetti2009community}, Infomap stands as one of the best
community detection algorithms in the literature. 

Another category of algorithms is formed by those capable of finding
overlapping communities, which have gained significant interest during the last
years.  We find several proposals, such as
Oslom~\cite{lancichinetti2011finding}, which uses the \emph{significance} as a
fitness measure in order to assess the quality of a community. Similar to
modularity, the significance is defined as the probability of finding a given
cluster in a random null model. Another algorithm that falls into this
category is the Link Clustering Algorithm (LCA)~\cite{ahn2010link}. This
algorithm is based on the idea of taking edges instead of vertices to form a
community.  The similarity of adjacent vertices is assessed by looking at the
Jaccard coefficient of the adjacency lists of the two vertices of the edges.
Those edges connecting vertices with high similarity are assigned to the same
community, and so overlapping communities emerge naturally. Finally, a recently
proposed
algorithm is BigClam by Yang et al.~\cite{YangL13}. This algorithm is based on
computing an affiliation of vertices to communities that maximizes an objective
function using non-negative matrix factorization. The objective function is
based on the intuition that the probability of an edge existing between two
vertices increases with the number of communities the vertices share (i.e. the
number of communities in which the vertices overlap).

Most of the work regarding the exploitation of parallelism for community
detection has the form of multithreaded algorithms for SMP machines.
In~\cite{lu2014parallel}, authors propose a parallel version of the
\emph{Louvain} method, which achieves an speedup of 16x using 32 threads.
Similarly, in~\cite{riedy2012scalable} Riedy et al. propose an agglomerative
modularity optimization algorithm for the Cray XMT and Intel based machines,
capable of analyzing a graph with 100 million nodes and 3.3 billion edges in
500 seconds.  Finally, in ~\cite{bae2013scalable} the authors propose a
parallel version of Infomap, called RelaxMap that relaxes concurrency
assumptions of the original method, achieving a parallel efficiency of about
70\%.

There has been little work regarding distributed algorithms for community
detection. One family of algorithms that fit well into the vertex-centric model
are those based on label propagation~\cite{raghavan2007,xie2012}.  In label
propagation, each vertex is initialized with a unique label, and then, they
define rules that simulate the spread of these labels in the network similarly
to infections. Label propagation has the advantage of being asymptotically
efficient, but no theoretical guarantees are given regarding the quality of the
results, especially in networks where communities are not well-defined.

\section{Background \& Terminology}
\label{sec:background}

Informally stated, the goal of community detection is, given a graph, to divide
the graph into groups (communities) of vertices such that, within a group,
vertices are tightly connected, and between groups, there are few connections.
For non-overlapping community detection, which is the focus of this paper, no
two communities contain the same vertex.  There are two primary aspects of
this problem. First, it is necessary to give a formal definition of a metric
that defines the quality of a given grouping, or {\em partitioning}, of a graph.
The next step is to create an algorithm to find one or more partitionings of the
graph that optimize this metric.

In this paper, we address the second part of this problem by proposing a
scalable, distributed algorithm for the optimization of the WCC metric proposed
in \cite{wcc}. This metric is defined on unweighted, undirected graphs. Inspired
by properties of real-life networks, the basic idea behind the metric is that
within a community, vertices should have a high concentration of triangles among
themselves, and they should close more triangles with other vertices in the
community than with vertices outside of the community. Using this idea, given an
undirected, unweighted graph $G(V, E)$, the quality of a community may then be
defined as the average cohesion of each of its member vertices to the other
vertices in the community, where the cohesion of a vertex $x$ to a set of
vertices $S$ is defined as 
\[
WCC(x, S) =
\begin{cases}
    \frac{t(x, S)}{t(x, V)}\cdot\frac{vt(x, V)}{|S \backslash \{x\}| + vt(x,
    V \backslash S)}& \text{if } t(x, V) \neq 0\\  
    0 & \text{if } t(x, V) = 0
\end{cases}
\]
The function $t(x, S)$ here gives the number of triangles closed by $x$ with
other vertices in $S$, and the function $vt(x, S)$ gives the number of unique
vertices contained in all such triangles. This cohesion metric therefore rewards
a high ratio of triangles closed within the community versus triangles closed
outside of the community (the left-hand term) and punishes vertices that have a
high number of vertices in its community with which it does not close any triangle
(the right-hand term). In other words, the left term promotes that the 
communities are well defined and isolated from the rest of the graph; 
and the right term promotes that all nodes in
the community are interconnected and form triangles.

The quality of a partitioning is the average
quality of each vertex in its assigned community. So, for a set
$S$, $WCC(S)$ is defined as the average $\forall x \in S$ of $WCC(x, S)$, and
the final $WCC$ of a partitioning $\mathcal{P} = \{C_1, \dots, C_n\}$ of $V$ is
then defined as 
\[
WCC(\mathcal{P}) = \frac{1}{|V|} \sum_{S \in \mathcal{P}} \sum_{x \in S} WCC(x, S).
\]
In practice, the optimization of this metric results in high quality
partitionings that correspond well to ground-truth communities, and it satisfies a
number of desirable theoretical properties. For more information on the metric
itself see \cite{wcc}. 

\section{Presentation of Algorithm} 
\label{sec:algorithm}

\begin{figure*}
\centerline{\includegraphics[width=\textwidth]{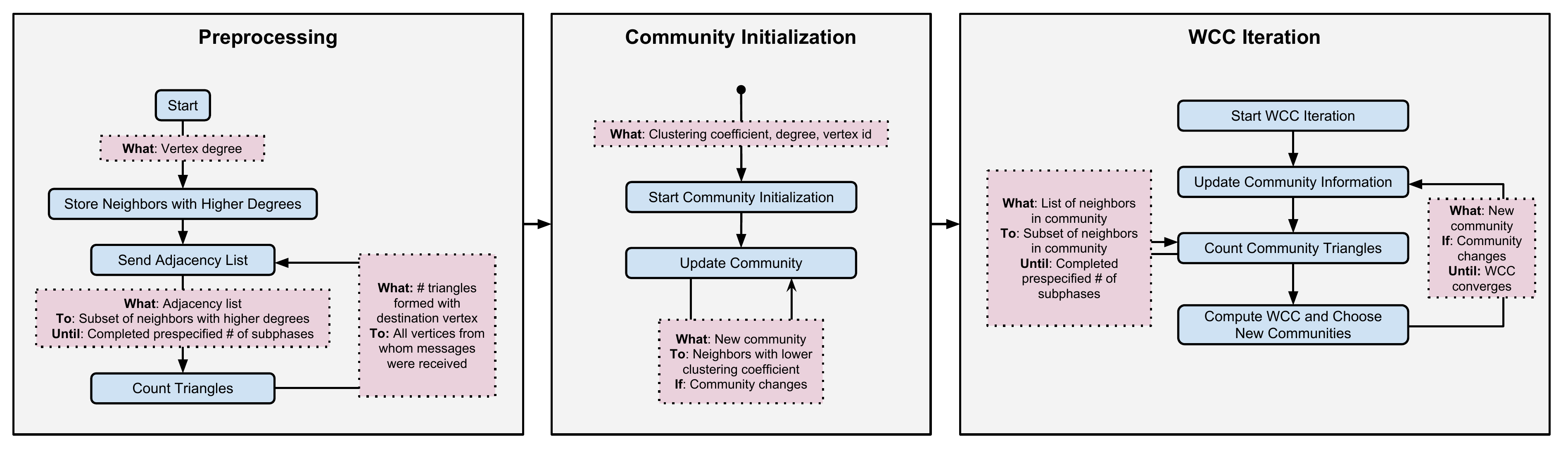}}
\caption{Algorithm Overview. The boxes with dotted edges represent messages sent
by vertices. Where unspecified, assume that a message is sent to all neighbors
of the vertex.}
\label{fig:alg_overview}
\end{figure*}

Our algorithm for optimizing 
this metric consists of three basic phases:
preprocessing, community initialization, and WCC iteration. In the first phase,
the values of $t(x, V)$ and $vt(x, V)$ are computed for every vertex, and all
edges that do not belong to any triangles are removed from the graph. Next, the
local clustering coefficient of each vertex is computed, and an initial
partitioning of the graph is determined based on these coefficients. From this
initial partitioning, the WCC iteration process is repeatedly applied, where
each vertex chooses a new community simultaneously based on a heuristic and the
global WCC value is computed. The algorithm halts when the WCC value converges.
An overview of the algorithm can be seen in Figure~\ref{fig:alg_overview}.

\subsection{Preprocessing}

The preprocessing portion of the algorithm is responsible for two things: counting,
for each vertex, the total of number of triangles it belongs to in the graph
($t(x, V)$), and removing all edges which do not belong to any triangles. After
removing all such edges, $vt(x, V)$ is simply the degree of the vertex $x$. This
filtering step improves performance and allows simplifying assumptions later
when deciding whether to transfer a vertex from one community to another. Note
that these two values are constant throughout computation and therefore only
need to be calculated once.

Given two vertices $u$ and $v$,  a standard way to compute the number of
triangles they form together (the number of triangles in which the edge $(u, v)$
is included) is to intersect their adjacency lists in order to count the number
of their common neighbors. If the two vertices have no common neighbors, the
edge $(u, v)$ is removed from the graph, because it does not affect the
computation of WCC. To count all of the triangles in the
graph in which node $u$ is contained, one must do this process for every
neighbor $v$ of $u$. In a centralized setting, this is relatively
straightforward to implement.  However, in a vertex-centric distributed setting,
vertices do not have access to the adjacency lists of their neighbors, and
therefore adjacency lists must be sent between vertices via message passing.
With a large graph, if every vertex sends its adjacency list to every one of its
neighbors in one superstep, this may lead to an excessive amount of time being
spent in communication, or in the worst case, to memory problems that
cause worker failures.

In order to address this problem, we propose two optimizations. First, we
observe that in real life graphs, there tend to be a few `hub' vertices 
with a very high degree and many vertices with a much lower 
degree~\cite{DBLP:conf/kdd/LeskovecKF05}. This
means that when these hub vertices send out their adjacency sets, it incurs a high
communication cost in comparison with the messages sent by non-hub vertices.
For this reason, in the first superstep, each vertex sends its degree to all of
its neighbors, and following this, vertices only send their adjacency sets to
neighbor vertices with a higher degree. The higher degree vertex in an edge then
counts the triangles formed with the lower degree vertex, and responds with a
message containing the triangle count. 

Secondly, to reduce the occurrence of memory problems, this phase may be split
into several subphases, where each vertex only sends its adjacency list to a
subset of its neighbors in each subphase. The number of subphases is chosen by
counting the total number of vertices that will be sent in messages during
preprocessing and using this to estimate the approximate overhead required to
send these messages, yielding the model
\[
nPrepPhases = \Bigg\lceil{\frac{vertexSize \cdot \sum_{v \in V} |adj(v)||hdn(v)|}
                      {nWorkers \cdot availWorkerMemory}}\Bigg\rceil,
\]
where $adj(v)$ is the adjacency set of vertex~$v$ (the contents of a
preprocessing message), $hdn(v)$ is the set of neighbors of $v$ that have a
higher degree than it (the destinations of the message), and $vertexSize$ is an
estimate of the amount of memory taken to send one vertex id. For a given
vertex~$v$, $|adj(v)||hdn(v)|$ gives the total number of elements
(vertex ids) that will be sent during preprocessing. The numerator therefore
estimates the total amount of memory that will be taken by all messages sent
across all preprocessing phases. This sum is computed with aggregators just
after the computation of $hdn$ for each vertex. The denominator estimates the
total amount of memory available for preprocessing overhead in the cluster,
assuming an even degree distribution across workers. The value for
$availWorkerMemory$ is chosen based on the resources available for
preprocessing, and the number of preprocessing phases is thus chosen such that
each subphase operates with an overhead less than this value.

Together, these two optimizations together greatly reduce the cost of
communication during preprocessing.

\subsection{Community Initialization}

Following preprocessing, the graph consists only of edges that are part of at
least one triangle. The next step is to create an initial partitioning of the
graph from which to begin the process of WCC optimization, meaning that each vertex
must decide its initial community. We make the
assumption that the higher the clustering coefficient of a vertex, the more
likely its neighbors are to belong to its community, because a high clustering
coefficient indicates that these vertices are tightly connected. 
This assumption is also applied in \cite{scd}, but the computation method
presented there is not adapted to the vertex centric processing model.

\begin{figure}
\centerline{\includegraphics[width=2.5in]{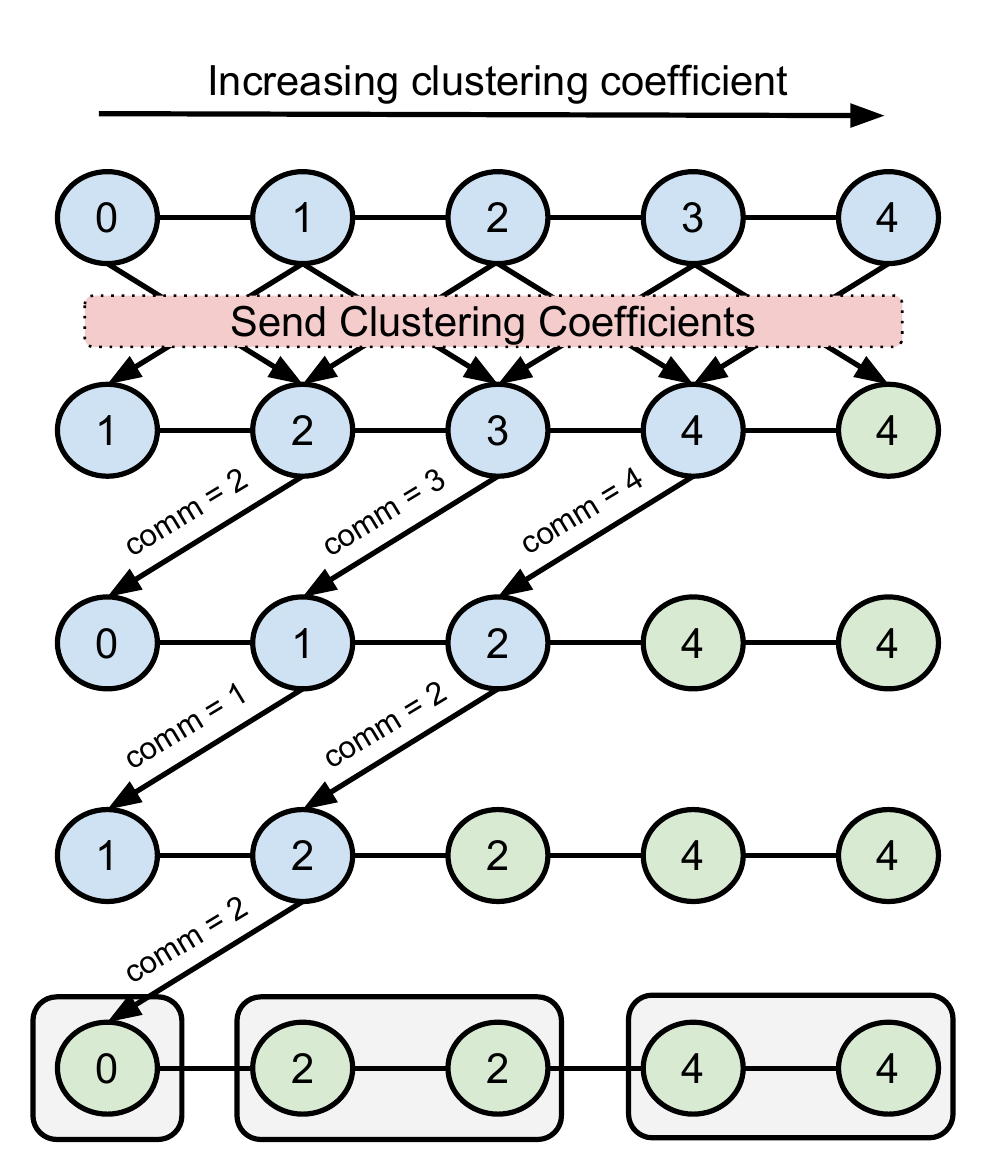}}
\caption{Community Initialization Process. The number inside each vertex
indicates its current community, and arrows represent messages being passed from
one vertex to another. Execution proceeds downward in the diagram, with each
level being the resulting configuration after the incoming messages are
processed. When a vertex receives a message from a vertex with id $x$ and
community $y$, it checks whether $x$ is currently a center vertex by checking if $x
= y$ and whether the clustering coefficient of $x$ is greater than that of its
current center and changes its community accordingly.}
\label{fig:comm_init}
\end{figure}

We require that the initial
communities fulfill the following properties:
\begin{enumerate}
\item Every community contains a single center vertex and a set of border vertices connected
to the center vertex.
\item The center vertex has the highest clustering coefficient of any vertex 
in the community.
\item Given a center vertex $y$ and a border vertex $x$ in a community,
the clustering coefficient of $y$ must be higher than the clustering
coefficient of any neighbor $z$ of $x$ that is the center of its own community.

\end{enumerate}
The process for obtaining such initial communities is shown
in Figure~\ref{fig:comm_init}. First, each vertex sends a message with
its id, its clustering coefficient, its degree\footnote{The degree is only used
in the case that two neighbors of a vertex have identical clustering
coefficients. In this case, the vertex with the higher degree is considered to
be `higher'. If the degrees are also equal, the vertex with the higher id is
considered to be higher.}, and its initial community (its vertex id) to all of
its neighbors. Each vertex then saves its incoming messages for use in future
steps. Following this step, a vertex chooses its new community to be the id of
the neighbor who has the highest clustering coefficient, considering as
candidates only the neighbors that are currently centers. If its
own clustering coefficient is higher then that of any neighbor or if none of its
neighbors are currently centers, it chooses to be the center of its own
community.  

In the example in the figure, this means that after the
communication of clustering coefficients, each vertex chooses the id of the
vertex to its right as its community. However, after this step, the third
desired property above is violated; the first three nodes belong to the
communities of the vertices to their right, none of which are center nodes. So,
it is then necessary for any vertex $x$ that has changed communities to
communicate its new community to all of its neighbors with lower
clustering coefficients. These neighbors are the only ones that need to be
notified because only vertices with a lower coefficient can become border nodes
of $x$.  After receiving the new communities of their neighbors, vertices
redetermine their communities based on which neighbors have become borders and
centers in the previous step. This iterative process continues until no vertices
change communities, in which case all three properties above are satisfied.

\subsection{WCC Iteration}

The main idea behind WCC iteration is to have each vertex repeatedly update its
community based on an improvement heuristic and to evaluate the overall WCC
between each update, and after a prespecified number of steps where the WCC does
not improve more than a certain amount, the computation halts. 
\subsubsection{Choosing a new community} 
When updating its community, the vertex has three options:
\begin{itemize}
\item {\bf Transfer:} The vertex moves from its community to the community of a
neighboring vertex.
\item {\bf Remove:} The vertex removes itself from its current community and
becomes the sole member of its own isolated community.
\item {\bf Stay:} The vertex remains in its current community.
\end{itemize}
In order to choose which of these actions to perform, the vertex must decide
which of the actions will most likely lead to the biggest improvement in the
global WCC value. In \cite{scd}, the authors present a heuristic for
the WCC improvement induced by each action, using aggregate community statistics for
a vertex's current and neighbor communities (the size and edge density of the
community and the number of edges leaving the community), the graph's clustering
coefficient, and a vertex's knowledge of its neighbors' communities. 
The heuristic is an approximation of the WCC that does not require the 
computation of the internal triangles, and thus is computationally more 
efficient.
Due to its effectiveness, we use this heuristic as well.  More details on the
heuristic can be found in \cite{scd}. Because this update process occurs
independently within each vertex, every vertex may perform the update
simultaneously, meaning that this portion of the algorithm very effectively
exploits parallelism. 

\subsubsection{WCC Computation}

In order to compute the actual global WCC, the values $t(x, C_x)$ and $vt(x, C_x)$
must be calculated for each vertex $x$ and its community $C_x$. This follows the
same distributed triangle-counting process as in preprocessing, except that
messages are only sent between vertices in the same community, and thus this
step is less computationally expensive than global triangle counting. These
local WCC values are then aggregated and averaged to obtain the global WCC. If a
new best WCC has been obtained, vertices save their current communities, and
when the WCC value converges, vertices output their saved community that led to
the best overall WCC.

\section{Experimentation}
\label{sec:experimentation}

For experimentation, we chose to perform tests on a variety of real life graphs,
%that have information about their ground truth communities, 
taken from the SNAP graph
repository\footnote{\texttt{http://snap.stanford.edu}}. Information
about each graph can be found in Table~\ref{tab:graphs}. Experiments
were performed on a 40 node cluster with 2.40GHz Xeon E5-2630L processors and
128G of RAM each, and a 1 Gigabit Ethernet connection between nodes. In terms of
software, we use Giraph release 1.1.0 and Hadoop version 0.20.203 (the default
for Giraph). %TODO: All experiments were repeated 3 times and their results were
%averaged. not true yet for friendster, still running...

\begin{table}[t]
\centering
%  \begin{tabular}{|c|c|c|c|c|c|c|}
\begin{tabular}{lrrr}
 \toprule
  & Vertices & Edges & Communities\\% & \% non-overlap & \% two-overlap & \% three-plus-overlap \\
 \midrule
%   \hline
  %Amazon & 334,863 & 925,872 & 151,037\\% & 3.9 & 3.6 & 92.4 \\
%   \hline
  Youtube       & 1,134,890 & 2,987,624 & 8,385\\% & 62.4 & 16.1 & 21.5 \\
%   \hline
  LiveJournal   & 3,997,962 & 34,681,189 & 287,512\\% & 35.8 & 17.2 & 47.0 \\
%   \hline
  Orkut         & 3,072,441 & 117,185,083 & 6,288,363\\% & 6.2 & 5.7 & 88.1\\
%   \hline
  Friendster    & 65,608,366 & 1,806,067,135 & 957,154\\% & 45.5 & 20.8 & 33.6\\
   \bottomrule
 \end{tabular}
\caption{Characteristics of the test graphs}
\label{tab:graphs}
\end{table}

\begin{figure}
\centering
\includegraphics[width=0.45\textwidth]{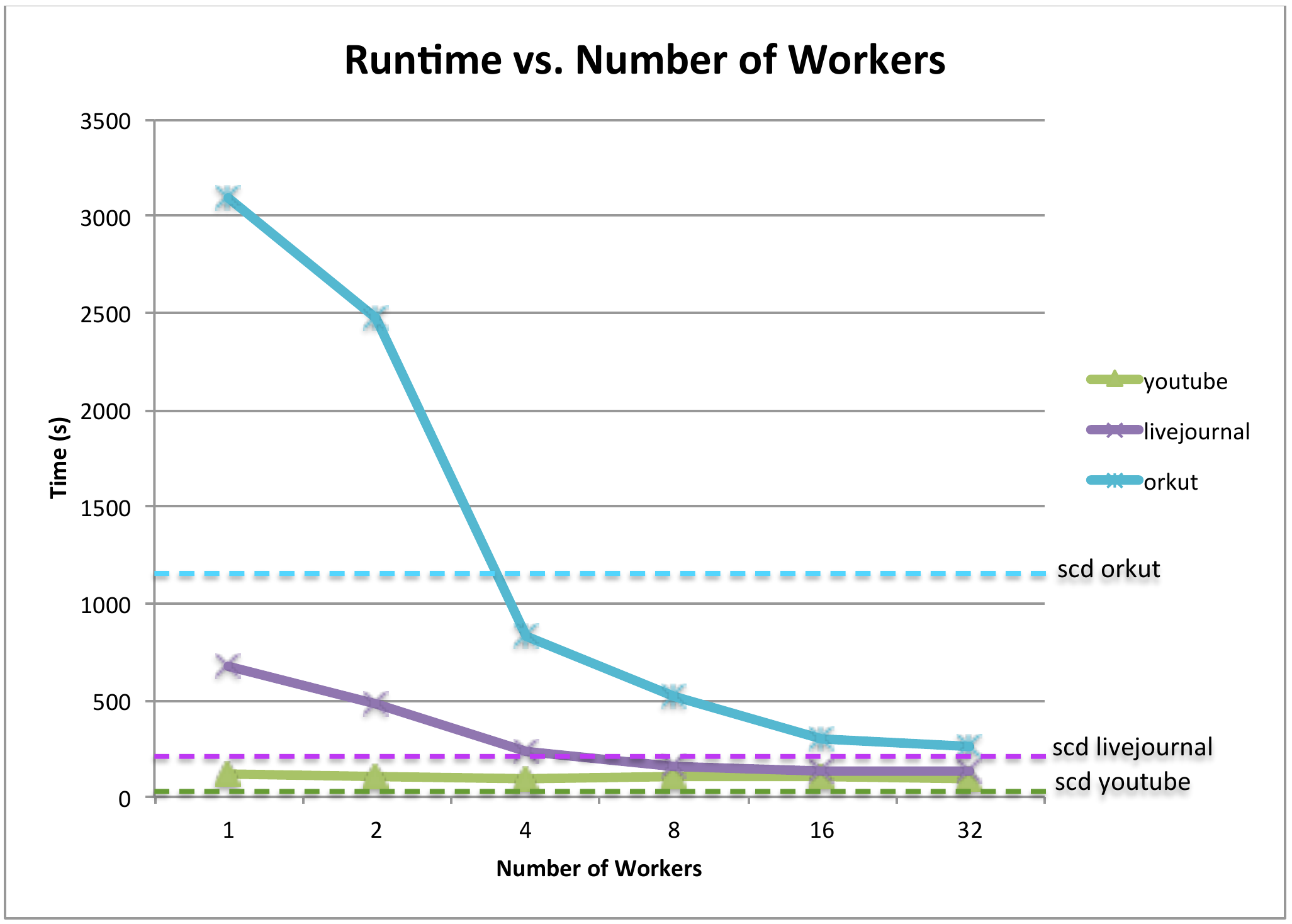}
\caption{The runtime of the algorithm on each graph, varying the number of
workers. The horizontal lines indicate the runtime of the centralized Scalable
Community Detection algorithm from \cite{scd}.}
\label{fig:runtime}
\end{figure}

\begin{figure}
\centering
\includegraphics[width=0.45\textwidth]{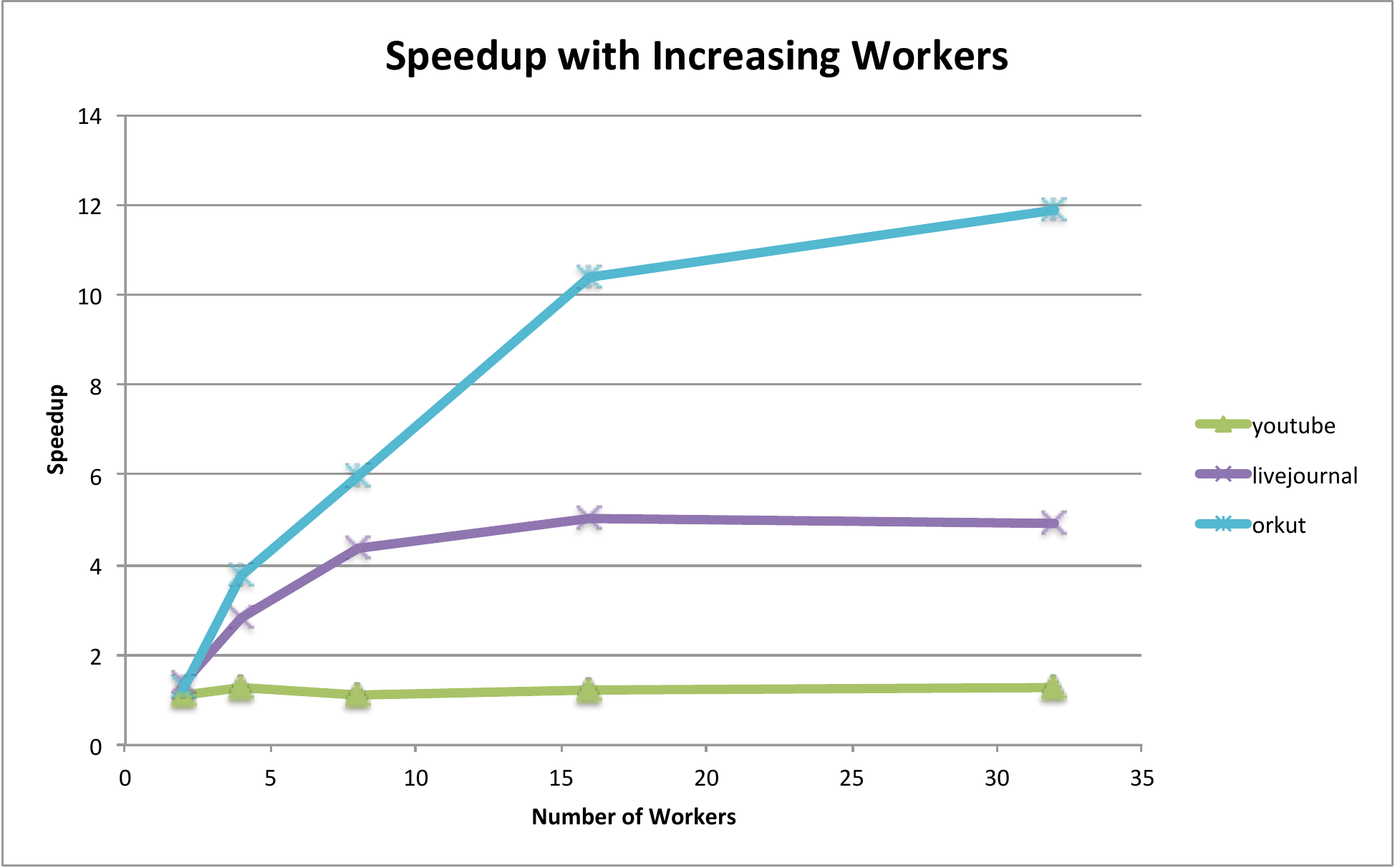}
\caption{The speedup of the algorithm on each graph as compared to the execution
time with 1 worker, varying the number of workers. Note that speedup improves
as the size of the graph increases.}
\label{fig:speedup}
\end{figure}

\begin{figure}
\centering
\includegraphics[width=0.45\textwidth]{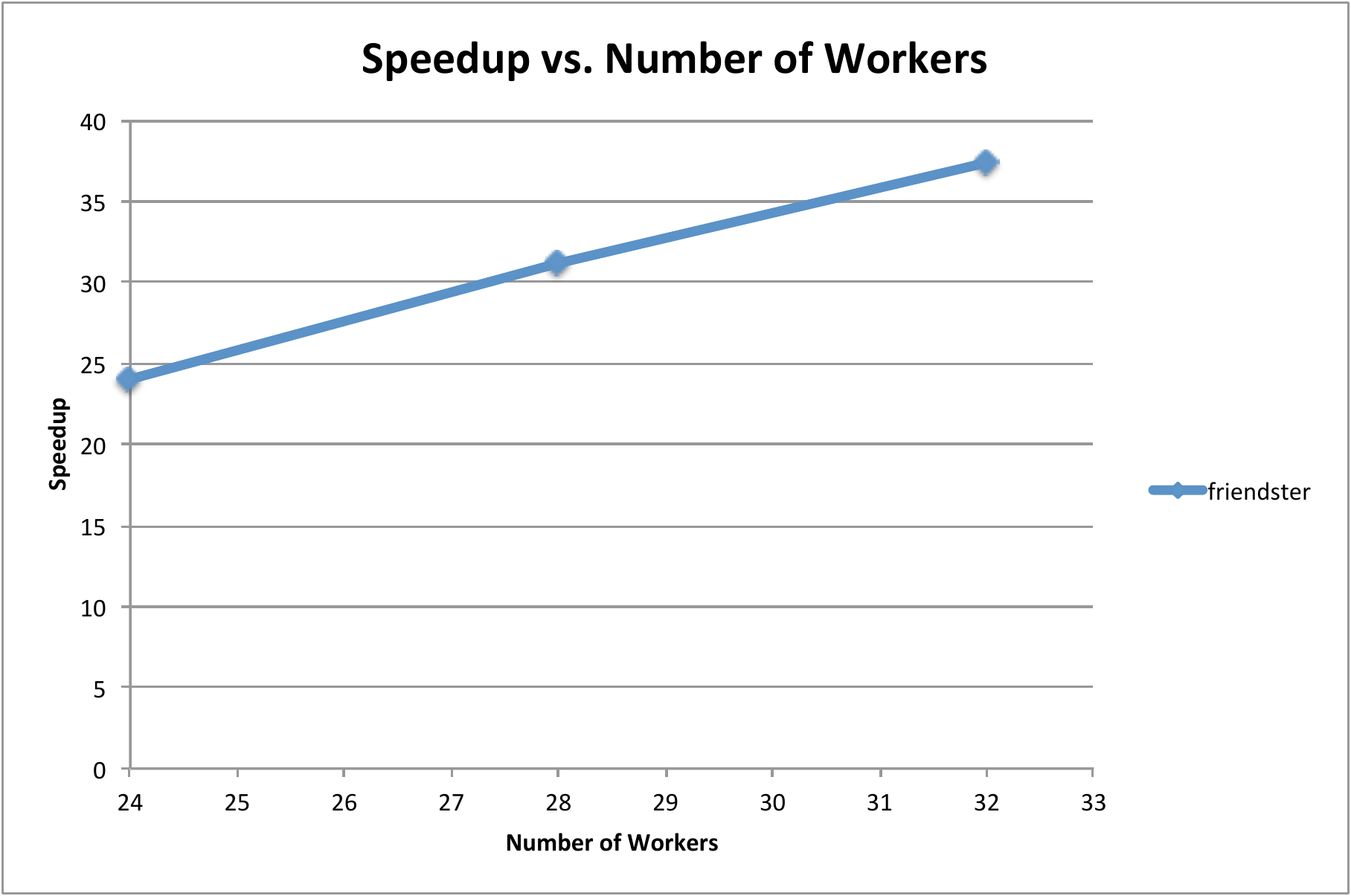}
\caption{The speedup of the Friendster graph, varying the number of workers.
Because the smallest number of machines used was 24, the runtime with one worker
is extrapolated to be 24 times the runtime with 24 workers, and the speedups
for the rest are calculated from that value. The runtime on 24 workers is
5826.234, much faster than the 29517.865 required by the centralized Scalable
Community Detection code.}
\label{fig:friendster_speedup}
\end{figure}

The goal of the experiments is to demonstrate the scalability of the method as
the number of workers grows, as well as to compare performance benefits and
result quality as compared to the parallel centralized version reported in
\cite{scd}. 
In Figure~\ref{fig:runtime}, we see that in all cases except
for the smallest graph, our distributed version eventually outperforms the
centralized version. The final WCC values obtained by both methods (not shown)
are very similar as well, indicating that there is no decrease 
in result quality incurred by the distributed algorithm. 
%TODO: @Matthew quantify the negligible quantity... less than 1 per cent? per million?...
In addition, from looking at 
Figures~\ref{fig:speedup} and \ref{fig:friendster_speedup}, the
speedup of the algorithm with the addition of workers improves with the size of
the graph; the larger the graph, the better the scalability. For the largest
graph, Friendster, we were not able to run the algorithm on fewer than 24
machines due to memory issues. 
This could be ameliorated by using a more efficient data
structure for storing the graph, since the graph alone used a large amount of
memory, which could be a topic of future work. In the largest graphs, we
measured that the two main bottlenecks are triangle counting during
preprocessing and in the computation of the next best community for each vertex.
Because the phase for choosing new communities is much more computation heavy
than communication heavy, it is to be expected that additional parallelism would
continue to boost performance especially in this phase as the number of workers
increases.  

\begin{figure}
\centering
\includegraphics[width=0.45\textwidth]{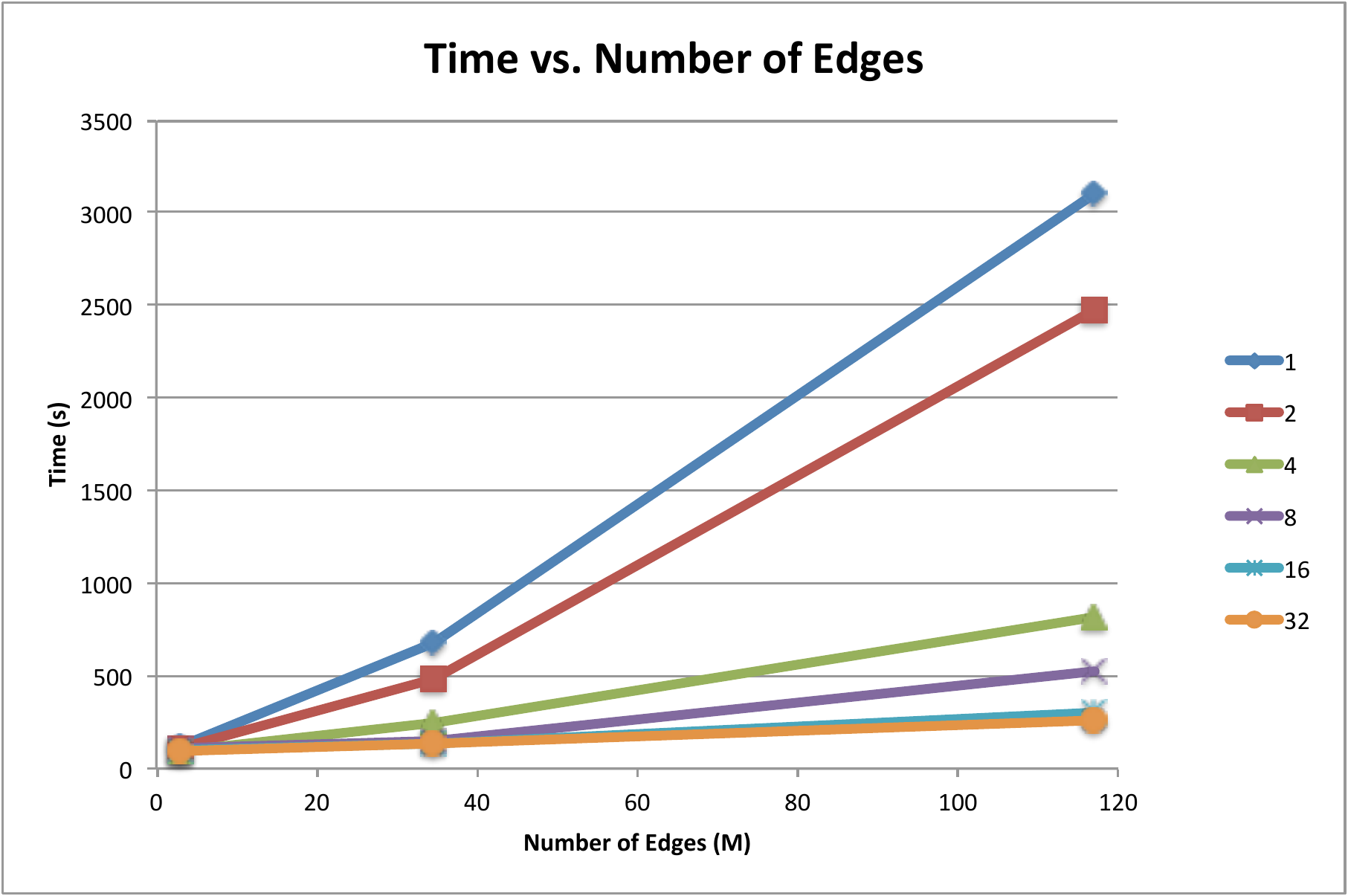}
\caption{The runtime of the algorithm increases with the number of edges in the
graph, varying the number of workers. When the number of workers is higher, the
runtime increases more slowly as edges are added.}
\label{fig:edges}
\end{figure}

\section{Conclusion \& Future Work} 
\label{sec:conclusion}
In this paper, we presented a scalable algorithm for distributed community
detection using the WCC metric that performs well on graphs of over one billion
edges. In particular, in the Friendster graph, with over 1.8 billion edges, 
the algorithm scales well from 24 to 32 workers and at its fastest finds all
communities in just one hour.  Current bottlenecks include memory
requirements for triangle counting and runtime for the computation of new
communities for each vertex. In the future, we may consider using alternative
triangle counting algorithms, including ones that only approximate the number
of triangles. Furthermore, implementing the algorithm in alternate frameworks
such as GraphX \cite{graphx} and GraphLab \cite{graphlab} would be worthwhile,
as well as comparing to other community detection algorithms like label
propagation.

\bibliographystyle{IEEEtran}
\bibliography{dist_wcc}
\end{document}